\documentclass[numberedappendix]{emulateapj}

\usepackage{apjfonts}

\usepackage{subfigure}

\newcommand{\ergps}{{\rm\,erg~s^{-1}}}
\newcommand{\msun}{{\rm M_{\odot}}}
\newcommand{\cm}{{\rm\,cm}}
\newcommand{\dF}{{^{^*}\!\!F}}

\begin{document}
\submitted{July 14, 2005}
\journalinfo{Web link for High Resolution document in footnote}

\title{Total and Jet Blandford-Znajek Power in Presence of Accretion Disk}

\author{Jonathan C. McKinney$^{1}$}
\altaffiltext{1}{Institute for Theory and Computation,
  Harvard-Smithsonian Center for Astrophysics, 60 Garden Street, MS
  51, Cambridge, MA 02138, USA\\
  High Res. Figures:
  \url{http://rainman.astro.uiuc.edu/\textasciitilde
    jon/bz.pdf}}
\email{jmckinney@cfa.harvard.edu}

\begin{abstract}

A rotating black hole probably powers ultrarelativistic jets in
gamma-ray bursts, relativistic jets from some active galactic nuclei,
and jets from some black hole x-ray binaries.  Prior estimates of the
power output of a black hole have assumed an infinitely thin disk, a
magnetic field based upon a slowly rotating black hole, and have not
self-consistently determined the geometry or magnitude of the magnetic
field for a realistic accretion disk.  We provide useful formulae for
the total and jet Blandford-Znajek (BZ) power and efficiency as
determined self-consistently from general relativistic
magnetohydrodynamic numerical models.  Of all jet mechanisms, we
suggest that only the BZ mechanism is able to produce an
ultrarelativistic jet.

\end{abstract}


\maketitle

\section{Introduction}\label{introduction}

For Poynting-dominated jets, where field lines tie the black hole to
large distances, the energy flux is determined by the Blandford-Znajek
(BZ) process \citep{bz77} (for a review see
\citealt{rw75,rees82,begelman84,bk00,punsly2001,mg04,lev05}).

The BZ effect depends on the magnetic field strength near the black
hole and the Kerr black hole spin parameter $a/M$, where $-1\le a/M\le
1$.  Self-consistent production of a relativistic Poynting jet likely
requires a rotating black hole accreting a thick disk with a disk
height ($H$) to radius ($R$) ratio of $H/R\gtrsim 0.1$
\citep{ga97,lop99,meier2001}.  As discussed below, a rapidly rotating
($a/M\sim 0.5-0.95$) black hole accreting a thick ($H/R\gtrsim 0.1$)
disk is probably common for jet systems.

However, prior estimates of the BZ power output assume the presence of
an infinitely thin ($H/R\sim 0$) disk, only apply for $a/M\sim 0$, and
do not self-consistently determine the magnetic field strength or
field geometry.  The purpose of this paper is to provide useful
formulae for the total and jet Blandford-Znajek power for arbitrarily
rapidly rotating black holes accreting a realistic disk.  We also
discuss the dominance of the BZ effect over other relativistic jet
mechanisms.

\section{Black Hole Accretion Systems}

The accretion of a thick disk around a rapidly rotating black hole is
often invoked as the engine to power GRBs
\citep{narayan1992,w93,pac98,mw99,narayan2001,brod04}.  Typical GRB
models invoke a relatively thick ($H/R\sim 0.1-0.9$) disk
\citep{mw99,pwf99,kohri2005}.  During the GRB event, the black hole
forms with $a/M\sim 0.5-0.75$ and evolves to a rapidly rotating state
with $a/M\sim 0.9-0.95$ \citep{narayan1992,mw99,shap02,shib02}.  GRB
models based upon internal shocks require an ultrarelativistic jet
with a typical Lorentz factor of $\Gamma\sim 100-1000$ in order to
overcome the compactness problem \citep{ls01,piran2005}, while Compton
drag models require $\Gamma\sim 20-100$
\citep{ghis00,lazzati2004,brod04}.  Direct observations of GRB
afterglow show evidence for relativistic motion
\citep{goodman1997,taylor2004a,taylor2004b}. Large Lorentz factors
require a relatively large jet energy flux, which could be BZ-driven
and Poynting-dominated rather than neutrino-annihilation-driven and
enthalpy-dominated
\citep{mr97,pwf99,dimat02,mckinney2005a,mckinney2005b}.  Core-collapse
of a rapidly rotating star leads to an inner disk with a strong
uniform (perhaps net) poloidal field.

The accretion of a relatively thick ($H/R\sim 0.9$) disk around a
rapidly rotating black hole is probably the engine that powers jets
from AGN and some black hole x-ray binaries.  Both radio-loud AGN and
x-ray binaries in the low-hard state show a correlation between radio
and x-ray emission, which is consistent with radio synchrotron
emission and hard x-ray emission generated from Comptonization through
a thick disk \citep{merloni2003}.  This suggests the disk is
geometrically thick when a system produces a jet, where the disk is
probably ADAF-like with $H/R\sim 0.9$ \citep{ny95}.

Based on Soltan-type arguments, AGN each probably harbor a rapidly
rotating ($a/M\sim 0.9-0.95$) black hole
\citep{up95,erz02,gsm04,shapiro2005}.  AGN are observed to have jets
with $\Gamma\lesssim 10$ \citep{up95,biretta99}, even $\Gamma\sim 30$
\citep{brs94,gc01,jorstad01}, while some observations imply
$\Gamma\lesssim 200$ \citep{ghis93,kraw02,kono03}.  For example, the
jet in M87 shows a large-scale opening angle of $10^\circ$ with
$\Gamma\sim 6$ \citep{junor99,biretta2002}. AGN probably accrete a
uniform field from solar wind capture or the ISM
\citep{narayan2003,pu05}.

Black hole x-ray binaries might have $a/M\sim 0.5-0.95$ \citep{gsm04},
while some may have $a/M\lesssim 0.5$ \citep{gd04}.  X-ray binary
systems produce outflows and jets \citep{mr99,mcclintock2003}.  For
example, black hole x-ray binary GRS 1915+105 has a jet with
apparently superluminal motion with $\Gamma\sim 1.5-3$
\citep{mr94,mr99,fb04,kaiser04}.  Solar-wind capture x-ray binaries
probably accrete a uniform field \citep{narayan2003}.

\section{The Blandford-Znajek Effect}\label{jets}

Most authors estimate the BZ power based upon the \citet{bz77} model
of a {\it slowly} spinning black hole threaded by a {\it
monopole}-based magnetic field and accreting an {\it infinitely thin
disk}, which gives
\begin{equation}\label{pbz}
P_{BZ,old} \approx P_0 \left(B^r[{\rm G}]\right)^2
\left(\Omega_H^2/c\right) r_g^4 ,
\end{equation}
where $B^r$ is the radial field strength, $r_g\equiv GM/c^2$,
$\Omega_H=ac/(2Mr_H)$ is the rotation frequency of the hole,
$r_H=r_g(1+\sqrt{1-(a/M)^2})$ is the radius of the horizon for angular
momentum $J=a GM/c$, and the dimensionless Kerr parameter has $-1\le
a/M\le 1$.  The parameter $P_0=0.01$ -- $0.1$, where the uncertainty
in $P_0$ arises because the strength of the magnetic field is not
self-consistently determined (see, e.g.,
\citealt{mt82,tm82,membranebook}).  Force-free numerical models agree
with the above BZ model \citep{kom01}. GRMHD numerical models of
slowly spinning, accreting black holes mostly agree with the BZ model
for the nearly force-free funnel region of the Poynting-dominated jet
\citep{mg04}.

The force-free solution for the monopole BZ flux is $\propto
\sin^2{\theta}$ \citep{bz77}, but the accretion of a relatively thick
disk diminishes total BZ power output substantially \citep{mg04}.
This is because the electromagnetic energy accreted as a disk
dominates the energy extracted.  Some black hole spin energy does
escape into the diffuse part of the corona, so the coronal outflow has
more Poynting flux for faster spinning holes
\citep{mg04,krolik05}. For rapidly rotating black holes, the field is
no longer monopolar and a significant amount of flux is generated
closer to the nearly force-free poles.

\begin{figure}
\includegraphics[width=3.33in,clip]{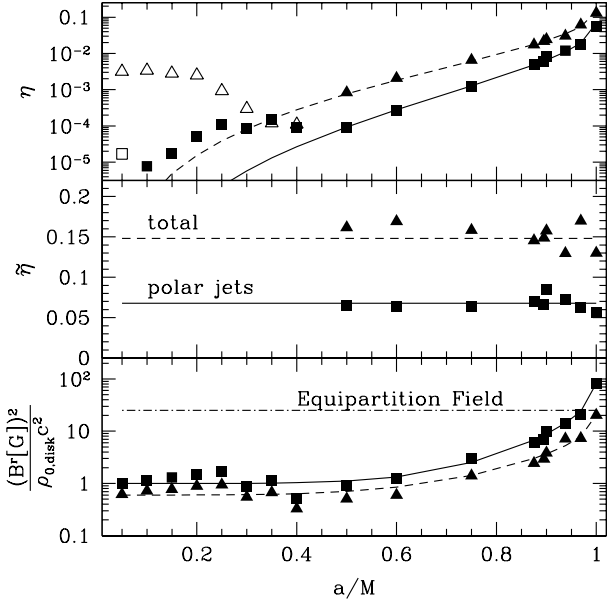}

\caption{Top panel: total (triangles: data, dashed line: fit) and jet
(squares: data, solid line: fit) efficiency.  Open points represent
negative efficiencies.  Middle panel: coefficient ($\tilde{\eta}$)
least squares fit to $\eta$ formulae.  Bottom panel: normalized field
(in Gauss) squared.}
\label{BZplot}
\end{figure}

HARM \citep{gmt03} was used to evolve a series of otherwise identical
GRMHD models with spin $a/M=-0.999 - 0.999$ and $H/R=0.1,0.2,0.5,0.9$
to determine the total BZ power ($P_{tot}$), jet BZ power ($P_{jet}$),
and field strength ($B^r$).  A similar series of models were studied
in \citet{mg04}, and they give a description of the model setup,
limitations, and related results.  The evolved field geometry is
relevant to most black hole systems and corresponds to a turbulent
disk field with a self-consistently generated large-scale flux
threading the black hole.

Figure~\ref{BZplot} shows the data and fits described below for
$H/R=0.2$.  There is a weak dependence on $H/R\gtrsim 0.1$ for the
{\it jet} BZ power or efficiency since the thicker the disk, the less
solid angle available to the jet, but the field strength there is
larger in compensation.  The non-jet results for other $H/R$ will be
presented in a separate paper.

The total (disk+corona+jet) BZ power efficiency in terms of the mass
accretion rate ($\dot{M}$) for $a/M>0.5$ is well fit by
\begin{equation}\label{EMTOTEFF}
\eta_{tot}=\frac{P_{tot}}{\dot{M}c^2} \approx 14.8\% \left(\frac{\Omega_H}{\Omega_H[a/M=1]}\right)^4 ,
\end{equation}
where the coefficient ($\tilde{\eta}$) is obtained by a least
squares fit.  Net electromagnetic energy is {\it accreted} for
$a/M\lesssim 0.4$ (including retrograde) when an accretion disk is
present.  This fact surprisingly agrees with the thin disk study of
\citet{li00}, who find that $\eta_{tot}>0$ only if $a/M\gtrsim
0.36$, corresponding to $\Omega_H>\Omega_K[{\rm ISCO}]$, the
Keplerian angular velocity at the inner-most stable circular orbit
(ISCO).  The sparse spin study of \citet{krolik05} is in basic
agreement with our $\eta_{tot}$.  Notice that the spin dependence of
our $\eta_{tot}$ is consistent with \citet{mg04} for their fit given
by equation 61 of data shown in figure 11, where they found
$\eta_{tot}\propto (2-r_H)^2\propto (a/M)^4$.  Their fit coefficient
of $6.8\%$ was inaccurate for $a/M\approx 1$ since they had no model
beyond $a/M=0.97$ and they included points with $a/M\lesssim 0.5$
that do not fit well.

The nearly force-free jet region contains field lines that tie the
black hole (not the disk) to large distances, leading to the BZ-effect
and a jet. For $a/M> 0.5$,
\begin{equation}\label{EMJETEFF}
\eta_{jet}=\frac{P_{jet}}{\dot{M}c^2} \approx 6.8\% \left(\frac{\Omega_H}{\Omega_H[a/M=1]}\right)^5
\end{equation}
over both polar jets.  If $a/M\approx 0.9$, then $\approx 1\%$ of the
accreted rest-mass energy is emitted back as a Poynting jet.

The horizon value of $B^r\equiv \dF^{rt}$, where $\dF$ is dual of the
Faraday, determines the black hole power output \citep{mg04}. For all
$a/M\ge 0$, the total and jet fields are
\begin{eqnarray}
\label{BRSQTOT}
\frac{(B^r_{tot}[{\rm G}])^2}{\rho_{0,disk}c^2} & \approx & 0.6+ 20\left(\frac{\Omega_H}{\Omega_H[a/M=1]}\right)^4,\\
\label{BRSQJET}
\frac{(B^r_{jet}[{\rm G}])^2}{\rho_{0,disk}c^2} & \approx & 1.0+ 81\left(\frac{\Omega_H}{\Omega_H[a/M=1]}\right)^5 ,
\end{eqnarray}
where the equipartition field satisfies $(B^r[{\rm
G}])^2/(8\pi)=\rho_{0,disk}c^2$, where $\rho_{0,disk}\equiv
\dot{M}t_g/r_g^3$ and $t_g\equiv GM/c^3$.  Hence,
\begin{eqnarray}
\label{PTOT}
P_{tot} & \approx & 7.4\times 10^{-3}((B^r_{tot}[{\rm G}])^2 r_g^2 c - 0.6\dot{M}c^2)
,\\\nonumber\\
\label{PJET}
P_{jet} & \approx & 8.4\times 10^{-4}((B^r_{jet}[{\rm G}])^2 r_g^2 c - 1.0\dot{M}c^2) ,
\end{eqnarray}
where since the field is determined self-consistently, no explicit
spin dependence appears.  This demonstrates the competition between
electromagnetic energy extraction and accretion.  Notice that no
direct comparison can be cleanly made to equation~\ref{pbz} due to the
presence of two ambiguities $P_0$ and $B^r$, while our formulae have
no ambiguities.  We suggest using the power output as given by
equations~\ref{EMTOTEFF} and~\ref{EMJETEFF}.  The black hole mass
accretion rate must be determined independent of $a/M$ using a
model-dependent study.

The coefficient in the formulae above depends on the type of accreted
field geometry.  As an extreme example, the accretion of a net
vertical field leads to an increase in the net electromagnetic
efficiency by a factor of five \citep{mg04}.  Also, the accretion of a
net toroidal field leads to negligible energy extraction \citep{dv05}.
Future studies should focus on the physical relevance, stability, and
long-temporal evolution of accreting net toroidal and vertical fields
with realistic perturbations rather than exact symmetries.  Accretion
of a net vertical field has been studied with nonrelativistic MHD
simulations \citep{igumenshchev2003}.

\section{BZ Jet Power for Collapsar Model}

For the collapsar model with black hole mass $M\sim 3\msun$ feeding at
an accretion rate of $\dot{M}=0.1\msun/s$, the magnetic field at the
poles of an $a/M\sim 0.9$ black hole is $B^r_{jet}\approx 10^{16}{\rm
G}$ and $\rho_{0,disk}\approx 3.4\times 10^{10}{\rm g}\cm^{-3}$.  This
gives a per polar axis Poynting flux jet energy of $P_{jet}\approx
10^{51}\ergps$. Notice that the neutrino annihilation jet luminosity
for such a collapsar model gives $L_{\nu\bar{\nu},ann,jet}\sim 10^{50}
- 10^{51}\ergps$ \citep{pwf99}, so these processes are likely both
important.  However, realistic models suggest that Poynting flux
dominates neutrino-annihilation energy flux
\citep{mckinney2005a,mckinney2005b}. Similar estimates can be made for
AGN and x-ray binaries.

\section{Dominance of Blandford-Znajek Effect}

\begin{figure}
\includegraphics[width=3.33in,clip]{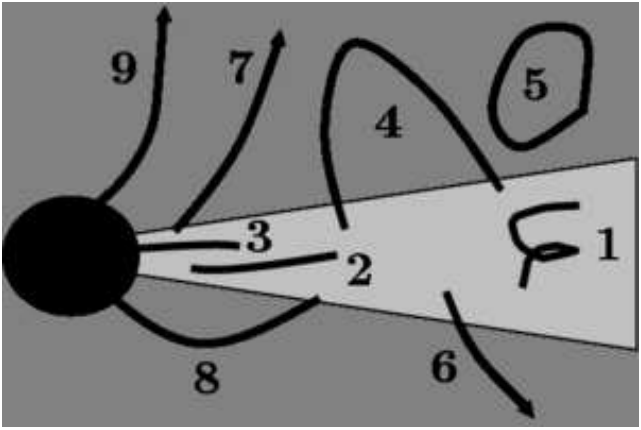}

\caption{Field type \#9 dominates in GRMHD numerical models and is
associated with Blandford-Znajek effect. Types \#1,2,3,5,6 dynamically
important.  Type \#4 transient. Types \#7,8 not dynamically
stable. }
\label{fieldtypes}
\end{figure}

Figure~\ref{fieldtypes} shows the possible types of field geometries
in the disk (see also, e.g., \citealt{blandford02,hirose04}).  Field
type \#1 corresponds to the Balbus-Hawley instability \citep{bh91},
which is present in our simulations.

Field type \#2 corresponds to models for which the field ties material
inside the ISCO to the outer disk \citep{gam99,krolik99}.  As they
predicted, unlike in the $\alpha$-viscosity model, there is no feature
at the ISCO or a direct plunge into the black hole
\citep{mg04,krolik05}, which impacts any radiative model of the
inner-radial accretion disk.  Field type \#3 corresponds to models
that consider the role of the black hole and disk on the disk
efficiency \citep{gam99,krolik99}.  They suggested that efficiencies
of order unity or higher could be achieved by extracting energy from
the hole.  This field type is present, but the disk efficiency is near
the thin disk efficiency \citep{mg04}.  Thus, surprisingly, the
magnetic field and disk thickness play little role in modifying the
disk efficiency.  In contrast, the angular momentum accreted is
reduced by magnetic field lines that tie from the disk to the hole
\citep{mg04,krolik05}.

Field types \#4 and \#5 correspond to surface reconnections.  Type \#4
geometries are temporary and type \#5 are common.  Thus, reconnection
efficiently removes large loops that tie the disk to itself.  Field
type \#6 corresponds to the Blandford-Payne type model \citep{bp82}.
We find that the lab-frame $|B|\propto r^{-5/4}$ as in their model,
but the lab-frame $\rho\propto r^{0.0}$ instead of their $\rho\propto
r^{-3/2}$ and there are few such field lines present in a stable
configuration due to the inner-radial corona being convectively
unstable and magnetically unstable to magnetic buoyancy.  Thin disks
likely have more stable surfaces that might allow for a stable wind.

Field type \#7 corresponds to coronal outflows or ergospheric-driven
winds \citep{pc90a,pc90b}.  There are no dynamically stable field
lines that tie the inner-radial disk to large distances.  Even for
$a/M=0.999$, no additional Poynting flux is created in the ergosphere
and the electromagnetic energy at infinity completely dominates the
hydrodynamic energy at infinity associated with the MHD Penrose
process, in basic agreement with the results of \citet{kom05} and
counter to the results of \citet{koide2002,punsly2005}.
\citet{koide2002} evolved for much too short a time.
\citet{punsly2005} used the 3D GRMHD near-horizon results of
\citet{krolik05}, but their near-horizon results could have numerical
artifacts associated with their use of Boyer-Lindquist coordinates.
However, there is a convectively and magnetically unstable,
self-consistently mass-loaded, collimated, mildly relativistic
($v/c\lesssim 0.95$) coronal outflow \citep{mg04,dv05}.  A rotating
black hole is not required for nonrelativistic ($v/c\lesssim 0.6$)
coronal outflows \citep{mg02,mg04}.

Field type \#8 corresponds to \citet{uzdensky2005} type models.  These
field geometries appear rarely and do not transfer a significant
amount of energy or angular momentum.  Such geometries may be more
important for thin disks.

Field type \#9, the dominant feature, is associated with the
Blandford-Znajek model.  Since the magnetic field confines the disk
matter away from the polar region, the rest-mass flux there is
arbitrarily low.  The large BZ flux to low rest-mass flux ratio can
translate into an arbitrarily fast jet.  The mass-loading of this jet
is considered in \citet{mckinney2005a,mckinney2005b}.

Notice that for accretion models with a net vertical field, the
resulting structure is essentially identical \citep{mg04}.
Reconnection efficiently erases the initial geometrical differences.

\section{Conclusions}

Typical BZ power output estimates assume an infinitely thin disk, a
slowly rotating black hole, and do not self-consistently determine the
magnitude or geometry of the magnetic field.  We use GRMHD numerical
models to self-consistently determine the total and jet BZ efficiency
when a disk is present.  There is a significantly stronger dependence
on black hole spin than prior estimates suggest.

Near the rotating black hole, the field geometry of the accretion
system is dominated by the field that leads to the BZ-effect.  Since
the polar region is magnetically confined against disk material and
the BZ power is large, the jet Lorentz factor can be arbitrarily
large.  This is unlike disk-related jet mechanisms that are directly
loaded by disk material.

\section*{Acknowledgments}

This research was supported by NASA-ATP grant NAG-10780 and a Harvard
CfA ITC fellowship.

\label{lastpage}


\begin{thebibliography}



%
%
%
\bibitem[Balbus \& Hawley(1991)]{bh91} Balbus, S.~A.~\& Hawley, J.~F.\
  1991, \apj, 376, 214
\bibitem[Begelman et al.(1984)]{begelman84} Begelman, M.~C.,
Blandford, R.~D., \& Rees, M.~J.\ 1984, Reviews of Modern Physics,
56, 255 
\bibitem[Begelman et al.(1994)]{brs94} Begelman, M.~C., Rees,
M.~J., \& Sikora, M.\ 1994, \apjl, 429, L57 
\bibitem[Beskin \& Kuznetsova(2000)]{bk00} Beskin, V.~S., \&
Kuznetsova, I.~V.\ 2000, Nuovo Cimento B Serie, 115, 795


\bibitem[Biretta et al.(1999)]{biretta99} Biretta, J.~A., Sparks,
W.~B., \& Macchetto, F.\ 1999, \apj, 520, 621 
\bibitem[Biretta et al.(2002)]{biretta2002} Biretta, J.~A., Junor,
W., \& Livio, M.\ 2002, New Astronomy Review, 46, 239 

\bibitem[Blandford(2002)]{blandford02} Blandford, R.~D.\ 2002,
Lighthouses of the Universe: The Most Luminous Celestial Objects and
Their Use for Cosmology Proceedings of the MPA/ESO/, p.~381, 381


\bibitem[Blandford \& Znajek(1977)]{bz77} Blandford, R.~D.~\& Znajek,
  R.~L.\ 1977, \mnras, 179, 433
\bibitem[Blandford \& Payne(1982)]{bp82} Blandford, R.~D.,
\& Payne, D.~G.\ 1982, \mnras, 199, 883 
\bibitem[Broderick(2004)]{brod04} Broderick, A.~E.\ 2004, astro-ph/0411778

\bibitem[De Villiers et al.(2005a)]{dv05} De Villiers, J.,
Hawley, J.~F., Krolik, J.~H., \& Hirose, S.\ 2005, \apj, 620, 878 
\bibitem[Di Matteo et al.(2002)]{dimat02} Di Matteo, T., Perna,
R., \& Narayan, R.\ 2002, \apj, 579, 706 
\bibitem[Elvis, Risaliti, \& Zamorani(2002)]{erz02} Elvis, M.,
        Risaliti, G., \& Zamorani, G.\ 2002, \apjl, 565, L75
\bibitem[Fender \& Belloni(2004)]{fb04} Fender, R., \&
Belloni, T.\ 2004, \araa, 42, 317 
\bibitem[Gammie(1999)]{gam99} Gammie, C.F. 1999, \apjl, 522, L57
%
\bibitem[Gammie~et~al.(2003a)]{gmt03} Gammie, C. F., McKinney, J. C.,
\& G$\acute{a}$bor T$\acute{o}$th 2003, \apj, 589, 444 
\bibitem[Gammie, Shapiro, \& McKinney(2004)]{gsm04} Gammie,
C.~F., Shapiro, S.~L., \& McKinney, J.~C.\ 2004, \apj, 602, 312 
\bibitem[Gierli{\' n}ski \& Done(2004)]{gd04} Gierli{\'
n}ski, M., \& Done, C.\ 2004, \mnras, 347, 885 
%
\bibitem[Ghisellini et al.(1993)]{ghis93} Ghisellini, G.,
Padovani, P., Celotti, A., \& Maraschi, L.\ 1993, \apj, 407, 65 
\bibitem[Ghisellini et al.(2000)]{ghis00} Ghisellini, G.,
Lazzati, D., Celotti, A., \& Rees, M.~J.\ 2000, \mnras, 316, L45 
\bibitem[Ghisellini \& Celotti(2001)]{gc01} Ghisellini, G.,
\& Celotti, A.\ 2001, \mnras, 327, 739 
\bibitem[Ghosh \& Abramowicz(1997)]{ga97} Ghosh, P.~\&
Abramowicz, M.~A.\ 1997, \mnras, 292, 887 
\bibitem[Goodman(1997)]{goodman1997} Goodman, J.\ 1997, New
Astronomy, 2, 449 
\bibitem[Hirose et al.(2004)]{hirose04} Hirose, S., Krolik,
J.~H., De Villiers, J., \& Hawley, J.~F.\ 2004, \apj, 606, 1083 
\bibitem[Igumenshchev et al.(2003)]{igumenshchev2003} Igumenshchev,
I.~V., Narayan, R., \& Abramowicz, M.~A.\ 2003, \apj, 592, 1042 
\bibitem[Jorstad et al.(2001)]{jorstad01} Jorstad, S.~G.,
Marscher, A.~P., Mattox, J.~R., Wehrle, A.~E., Bloom, S.~D., \&
Yurchenko, A.~V.\ 2001, \apjs, 134, 181 
\bibitem[Junor et al.(1999)]{junor99} Junor, W., Biretta,
J.~A., \& Livio, M.\ 1999, \nat, 401, 891 
\bibitem[Kaiser et al.(2004)]{kaiser04} Kaiser, C.~R., Gunn,
K.~F., Brocksopp, C., \& Sokoloski, J.~L.\ 2004, \apj, 612, 332 
\bibitem[Kohri et al.(2005)]{kohri2005} Kohri, K., Narayan, R.,
\& Piran, T.\ 2005, astro-ph/0502470 
\bibitem[Koide et al.(2002)]{koide2002} Koide,
S., Shibata, K., Kudoh, T., \& Meier, D.~L.\ 2002, Science, 295,
1688 
\bibitem[Komissarov(2001)]{kom01} Komissarov, S.~S.\ 2001,
\mnras, 326, L41

\bibitem[Komissarov(2005)]{kom05} Komissarov, S.~S.\ 2005,
\mnras, 359, 801

\bibitem[Konopelko et al.(2003)]{kono03} Konopelko, A.,
Mastichiadis, A., Kirk, J., de Jager, O.~C., \& Stecker, F.~W.\
2003, \apj, 597, 851 
\bibitem[Krawczynski et al.(2002)]{kraw02} Krawczynski, H.,
Coppi, P.~S., \& Aharonian, F.\ 2002, \mnras, 336, 721 
\bibitem[Krolik(1999)]{krolik99} Krolik, J.~H.\ 1999, \apjl,
515, L73 

\bibitem[Krolik et al.(2005)]{krolik05} Krolik, J.~H., Hawley,
J.~F., \& Hirose, S.\ 2005, \apj, 622, 1008


\bibitem[Lazzati et al.(2004)]{lazzati2004} Lazzati, D., Rossi, E.,
Ghisellini, G., \& Rees, M.~J.\ 2004, \mnras, 347, L1 
\bibitem[Levinson(2005)]{lev05} Levinson, A.\ 2005, astro-ph/0502346


\bibitem[Li(2000)]{li00} Li, L.\ 2000, \apjl, 533, L115


\bibitem[Lithwick \& Sari(2001)]{ls01} Lithwick, Y., \&
Sari, R.\ 2001, \apj, 555, 540 
\bibitem[Livio, Ogilvie, \& Pringle(1999)]{lop99} Livio, M.,
Ogilvie, G.~I., \& Pringle, J.~E.\ 1999, \apj, 512, 100 
\bibitem[MacDonald \& Thorne(1982b)]{mt82} MacDonald, D.~\&
Thorne, K.~S.\ 1982, \mnras, 198, 345 
\bibitem[MacFadyen \& Woosley(1999)]{mw99} MacFadyen, A. I. \& Woosley, S. E.\
1999, \apj, 524, 262 
%
\bibitem[McClintock \& Remillard(2003)]{mcclintock2003} McClintock,
J.~E., \& Remillard, R.~A.\ 2003, astro-ph/0306213

\bibitem[McKinney \& Gammie(2002)]{mg02} McKinney, J.~C., \& Gammie,
C.~F.\ 2002, \apj, 573, 728 
\bibitem[McKinney \& Gammie(2004)]{mg04} McKinney, J.~C., \&
Gammie, C.~F.\ 2004, \apj, 611, 977 
%
\bibitem[McKinney(2005a)]{mckinney2005a} McKinney, J.~C. 2005a, astro-ph/0506368

\bibitem[McKinney(2005b)]{mckinney2005b} McKinney, J.~C. 2005b, astro-ph/0506369

\bibitem[Meier(2001)]{meier2001} Meier, D.~L.\ 2001, \apjl, 548, L9
\bibitem[Merloni et al.(2003)]{merloni2003} Merloni, A., Heinz, S.,
\& di Matteo, T.\ 2003, \mnras, 345, 1057 
\bibitem[M{\' e}sz{\' a}ros \& Rees(1997)]{mr97} M{\' e}sz{\' a}ros, P.~\& Rees,M.~J.\ 1997, \apjl, 482, L29
\bibitem[Mirabel \& Rodriguez(1994)]{mr94} Mirabel, I.~F.,
\& Rodriguez, L.~F.\ 1994, \nat, 371, 46 
\bibitem[Mirabel \& Rodr{\'{\i}}guez(1999)]{mr99} Mirabel,
I.~F., \& Rodr{\'{\i}}guez, L.~F.\ 1999, \araa, 37, 409 
\bibitem[Narayan et al.(1992)]{narayan1992} Narayan, R., Paczynski,
B., \& Piran, T.\ 1992, \apjl, 395, L83 
\bibitem[Narayan \& Yi(1995)]{ny95} Narayan, R.~\& Yi, I.\
1995, \apj, 452, 710 
\bibitem[Narayan et al.(2001)]{narayan2001} Narayan, R., Piran, T.,
\& Kumar, P.\ 2001, \apj, 557, 949 
\bibitem[Narayan et al.(2003)]{narayan2003} Narayan, R.,
Igumenshchev, I.~V., \& Abramowicz, M.~A.\ 2003, \pasj, 55, L69 
\bibitem[Paczynski(1998)]{pac98} Paczynski, B.\ 1998, \apjl,
494, L45 
\bibitem[Piran(2005)]{piran2005} Piran, T.\ 2005, Reviews of
Modern Physics, 76, 1143 
\bibitem[Popham et al.(1999)]{pwf99} Popham, R., Woosley,
S.~E., \& Fryer, C.\ 1999, \apj, 518, 356 
\bibitem[Punsly \& Coroniti(1990a)]{pc90a} Punsly, B., \&
Coroniti, F.~V.\ 1990a, \apj, 354, 583

\bibitem[Punsly \& Coroniti(1990b)]{pc90b} Punsly, B., \&
Coroniti, F.~V.\ 1990b, \apj, 350, 518



\bibitem[Punsly(2001)]{punsly2001} Punsly, B.\ 2001, Black hole
gravitohydromagnetics, Berlin: Springer, 2001, xii, 400 p.,
Astronomy and astrophysics library, ISBN 3540414665, 

\bibitem[Punsly(2005)]{punsly2005} Punsly, B.\ 2005, astro-ph/0505083


\bibitem[Rees et al.(1982)]{rees82} Rees, M.~J., Phinney,
E.~S., Begelman, M.~C., \& Blandford, R.~D.\ 1982, \nat, 295, 17 
\bibitem[Ruffini \& Wilson(1975)]{rw75} Ruffini, R., \&
Wilson, J.~R.\ 1975, \prd, 12, 2959 
\bibitem[Shapiro \& Shibata(2002)]{shap02} Shapiro, S.~L.~\& Shibata, M.\
2002, \apj, 577, 904 
\bibitem[Shibata \& Shapiro(2002)]{shib02} Shibata, M.~\& Shapiro, S.~L.\
2002, \apjl, 572, L39 
\bibitem[Shapiro(2005)]{shapiro2005} Shapiro, S.~L.\ 2005, \apj,
620, 59 
\bibitem[Taylor et al.(2004a)]{taylor2004a} Taylor, G.~B., Frail,
D.~A., Berger, E., \& Kulkarni, S.~R.\ 2004, \apjl, 609, L1 
\bibitem[Taylor et al.(2004b)]{taylor2004b} Taylor, G.~B., Momjian,
E., Pihlstrom, Y., Ghosh, T., \& Salter, C.\ 2004, astro-ph/0412483

\bibitem[Thorne \& MacDonald(1982)]{tm82} Thorne, K.~S., \&
MacDonald, D.\ 1982, \mnras, 198, 339 
\bibitem[Thorne, Price, \& MacDonald(1986)]{membranebook} Thorne,
K.~S., Price, R.~H., \& MacDonald, D.~A.\ 1986, Black Holes: The
Membrane Paradigm 
\bibitem[Urry \& Padovani(1995)]{up95} Urry, C.~M., \&
Padovani, P.\ 1995, \pasp, 107, 803 
\bibitem[Uzdensky(2005)]{uzdensky2005} Uzdensky, D.~A.\ 2005, \apj,
620, 889 
\bibitem[Uzdensky \& Spruit(2005)]{pu05} Uzdensky, D.~A., \&
Spruit, H.~C.\ 2005, astro-ph/0504429

\bibitem[Woosley(1993)]{w93} Woosley, S.~E.\ 1993, \apj,
405, 273 
%


\end{thebibliography}
\end{document}